\documentclass[twocolumn,english,aps,amsmath,showpacs,superscriptaddress]{revtex4}
\usepackage[T1]{fontenc}
\usepackage[latin1]{inputenc}
\usepackage{babel}
\usepackage{graphicx}
\usepackage{amsmath,amssymb,amsfonts,color}
\makeatletter

\providecommand{\LyX}{L\kern-.1667em\lower.25em\hbox{Y}\kern-.125emX\@}

\newcommand{\comment}[1]  {  }

\def\BE{\begin{equation}}
\def\EE{\end{equation}}
\def\BEA{\begin{eqnarray}}
\def\EEA{\end{eqnarray}}

\newcommand{\fd}[2]{\frac{d #1}{d #2}}
\newcommand{\pd}[2]{\frac{\partial #1}{\partial #2}}

\newcommand{\snr}{\textsf{snr}}

\newcommand{\ie}{\textsl{i.e.}}
\newcommand{\eg}{\textsl{e.g.}}

\makeatother
\begin{document}

\title{Shannon Meets Carnot: Generalized Second Thermodynamic Law}

\author {Ori Shental}
\affiliation{Center for Magnetic Recording Research, University of
California, San Diego\\9500 Gilman Drive, La Jolla, CA 92093, USA}

\author{Ido Kanter}
\affiliation{Department of Physics, Bar-Ilan University, Ramat-Gan,
52900 Israel}

\begin{abstract}
The classical thermodynamic laws fail to capture the behavior of
systems with energy Hamiltonian which is an explicit function of the
temperature. Such Hamiltonian arises, for example, in modeling
information processing systems, like communication channels, as
thermal systems. Here we generalize the second thermodynamic law to
encompass systems with temperature-dependent energy levels,
\mbox{$dQ=TdS+<d\mathcal{E}/dT>dT$}, where \mbox{$<\cdot>$} denotes
averaging over the Boltzmann distribution and reveal a new
definition to the basic notion of temperature. This generalization
enables to express, for instance, the mutual information of the
Gaussian channel as a consequence  of the fundamental laws of nature
- the laws of thermodynamics.
\end{abstract}


\maketitle

The current scientific conception is that the theory of information
is a creature of mathematics and has its own vitality independent of
the physical laws of nature.

The laws of thermodynamics are fundamental laws of nature, and in
particular, the second thermodynamic law linearly relates the change
in the entropy, \mbox{$dS$}, to the amount of heat, \mbox{$dQ$},
absorbed by a system at equilibrium, \mbox{$dQ=TdS$}, thus defining
the temperature, \mbox{$T$}, of the system~\cite{BibDB:Thermo}.
Thermodynamics is primarily an intellectual achievement of the
\mbox{$19^{th}$} century. The first analysis of heat engines was
given by the French engineer Sadi Carnot in his seminal 1824
publication, `\textit{on the Motive Power of Fire and on Machines
Fitted to Develop that Power}'~\cite{BibDB:Carnot,BibDB:Carnot2},
laying the foundations to the second law of thermodynamics. This
paper marks the start of thermodynamics as a modern
science~\cite{BibDB:HistoryOfThermo}, which was subsequently evolved
to the more general discipline of statistical
mechanics~\cite{BibDB:StatPhys}.

Information theory is a statistical theory dealing with the limits
and efficiency of informatics~\cite{BibDB:BookCover}. This theory is
one of the major enablers of modern technologies in the information
age, from compressed ZIP files, CDs, MP3s, DSL high-speed modems and
mobile phones to the Voyager missions to deep space. Similar to the
key role of Carnot's paper in the development of thermodynamics, the
birth of information theory as an independent discipline is
attributed to the landmark publication of Claude Shannon  in 1948
`\textit A Mathematical Theory of
Communication'~\cite{BibDB:Shannon}.

The generic problem in information processing is the transmission of
information over a noisy channel. This central paradigm of
information theory can be mathematically abstracted to having two
random variables \mbox{$X$} and \mbox{$Y$} representing the desired
information and its noisy replica, respectively. Noisy transmission
can occur either via space from one geographical point to another,
as happens in communications, or in time, for example, when writing
or reading files from a hard disk in the computer.

The most important measures of information are entropy and mutual
information. Information (Shannon) entropy, \mbox{$H(\cdot)$}, is a measure
of the amount of uncertainty in a random variable, indicating how
easily data can be compressed. Mutual information quantifies the
amount of information in common between two random variables and it
is used to upper bound the attainable rate of information
transferred across a channel. To put differently, mutual
information, \mbox{$I(X;Y)$}, measures the amount of information that can
be obtained about one random variable (the channel input \mbox{$X$}) by
observing another (the output \mbox{$Y$}). A basic property of the mutual
information is that \mbox{$I(X;Y)=H(X)-H(X|Y)$}, hence knowing \mbox{$Y$}, we can
save an average of \mbox{$I(X;Y)$} bits in encoding \mbox{$X$} compared to not
knowing \mbox{$Y$}.

The archetypal Gaussian channel is one of the most popular models in
the field of informatics and it arises in numerous applications in
information processing, modeling the relation between the latent
input and the corrupted observed output~\cite{BibDB:BookGallager2008}.
Consider a real-valued channel with input and output random
variables \mbox{$X$} and \mbox{$Y$}, respectively, of the form
\begin{equation}\label{eq_channel}
Y=X+N,
\end{equation}
where \mbox{$N\sim\mathcal{N}(0,1/\snr)$} is a Gaussian noise independent
of \mbox{$X$}, and \mbox{$\snr\geq0$} is the channel's signal-to-noise ratio
(SNR). The input is taken from a probability distribution \mbox{$P(X)$}
with a bounded second moment.
In this Letter, random variables are denoted
by upper case letters and their values denoted by lower case
letters.

The Gaussian channel~(\ref{eq_channel}) can be also viewed as a
physical system~\cite{BibDB:Sourlas,bibDB:Rujan,BibDB:BookNishimori}, operating under the laws of
thermodynamics. The microstates of the thermal
system are equivalent to the hidden values of the input \mbox{$X$}. A
comparison of the channel's a-posteriori probability distribution,
given by Bayes' law
\begin{eqnarray}
&&P{(X=x|Y=y)}=\frac{\sqrt{\snr}P(X=x)}{\sqrt{2\pi}p(Y=y)}\exp{\Big(-\frac{\snr}{2
}(y-x)^{2}\Big)}\nonumber\\&&=\frac{\exp{\big(-\snr(-xy+\frac{x^{2}}{2}-\frac{\log
P(X=x)}{\snr})\big)}}{\sum_{x\in\mathcal{X}}\exp{\big(-\snr(-xy+\frac{x^{2}}{2}-\frac{\log
P(X=x)}{\snr})\big)}},\nonumber
\end{eqnarray}
with the Boltzmann distribution law
yields the following mapping of the inverse temperature and energy
of the equivalent thermal system
\begin{eqnarray}\label{eq_snr_map}
    \snr&\rightarrow&\beta,\nonumber\\
    -xy+\frac{x^{2}}{2}-\frac{\log P(X=x)}{\beta}&\rightarrow&
\mathcal{E}(X=x|Y=y;\beta).\label{eq_energy_map}
\end{eqnarray}

In order to capture the temperature-dependent nature of the energy
in systems like the communication channel, we generalize the
formulation of the second law of thermodynamics \mbox{$dS=dQ/T$}.
The differential of the partition function's logarithm,
\mbox{$\log\mathcal{Z}$}, can be written as \BE
d\log\mathcal{Z}=\fd{\log\mathcal{Z}}{\beta}d\beta.\label{eq_identity}\nonumber\EE
Utilizing the identity \mbox{$\log\mathcal{Z}=-\beta U+S$}, one
obtains \BE dS=d(\beta U)+\fd{\log\mathcal{Z}}{\beta}d\beta.
\nonumber\EE Since for \mbox{$T$}-dependent energy \BE
\fd{\log\mathcal{Z}}{\beta}=-U-\beta\bigg<\fd{\mathcal{E}(X)}{\beta}\bigg>,\nonumber
\EE we get \begin{eqnarray} dS&=&d(\beta
U)-Ud\beta-\beta\bigg<\fd{\mathcal{E}(X)}{\beta}\bigg>d\beta\nonumber\\&=&\beta
dU-\beta\bigg<\fd{\mathcal{E}(X)}{\beta}\bigg>d\beta.\nonumber
\end{eqnarray} Recalling that according to the first law
\mbox{$dU=dQ$} concludes the generalized second thermodynamic law
\BE\label{eq_G_second_law}
dS=\frac{dQ}{T}-\frac{1}{T}\bigg<\fd{\mathcal{E}(X)}{T}\bigg>dT.\EE

The generalized second law of thermodynamics~(\ref{eq_G_second_law})
has a clear physical interpretation. For simplicity, let us assume
that an examined system is characterized by a comb of discrete
energy levels \mbox{$\mathcal{E}1,\mathcal{E}2,\ldots$} . The heat
absorbed into the \mbox{$T$}-dependent system has the following dual
effect: A first contribution of the heat,
\mbox{$dU-<d\mathcal{E}(X)/dT>dT$}, increases the temperature of the
system while the second contribution,
\mbox{$<d\mathcal{E}(X)/dT>dT$}, goes for shifting the energy comb.
However, the shift of the energy comb does \emph{not} affect the
entropy, since the occupation of each energy level remains the same,
and the entropy is independent of the energy values which stand
behind the labels \mbox{$\mathcal{E}1,\mathcal{E}2,\ldots$}. The
change in the entropy can be done only by moving part of the
occupation of one tooth of the energy comb to the neighboring teeth,
and this can be achieved only by changing the temperature. Hence,
the {\it effective heat} contributing to the entropy is
\mbox{$dQ-<d\mathcal{E}(X)/dT>dT$}, and this is the physical
explanation to the generalized second law~(\ref{eq_G_second_law}). A
schematic picture of communication-heat-engine is depicted in Figure
1, where the heat devoted to the change of the Hamiltonian is
denoted by the term `working channel'.

\begin{figure}
{{\includegraphics[angle=270,scale=0.3]{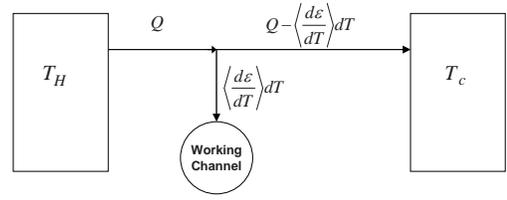}}
\par}
\caption{A schematic communication-heat-engine. H/C denote Hot/Cold temperatures. \label{enegine}}
\end{figure}

Note that the first law remains unaffected, \mbox{$dU=dQ$}, since both
ways of heat flow absorption into the system are eventually
contributing to the average internal energy \mbox{$U$}. The generalized
second law specifies the trajectory, the weight of each one of the
two possible heat flows, at a given temperature.

The temperature, originally defined by the second law as
\mbox{$T=dQ/dS$}, is redefined now as\BE\label{efficiency}
\frac{1}{T}=\frac{dS/dQ}{1-\frac{<d\mathcal{E}(X)/dT>}{dQ/dT}}=
\frac{dS/dQ}{1-\frac{<d\mathcal{E}(X)/dT>}{C_{V}(T)}}. \EE This
redefinition has a more complex form and involves an implicit
function of \mbox{$T$}, since the temperature appears on both sides
of the equation. Note that the form of the denominator
of~(\ref{efficiency}) resembles an efficiency factor of the
communication-heat-engine, and will be discussed at the concluding
paragraphs.

Based on the generalized second law, a thermodynamic
expression for the mutual information of quasi-static communication
channels can be derived. To make the derivation intelligible we
start with the thermodynamic expression for the mutual information in
the case of a temperature-independent Hamiltonian.

The entropy integral obtained from combining the three laws of thermodynamics for
the thermal system being equivalent to the Gaussian channel, is
\BEA\label{eq_S_final} S(\beta)=S(X|Y=y;\beta)
=-\int_{\beta}^{\infty}\gamma\frac{dU(Y=y;\gamma)}{d\gamma}d\gamma,\nonumber
\EEA where the internal energy,
\mbox{$U(Y=y;\beta)=\mathbb{E}_{X|Y}\{\mathcal{E}(X|Y=y)\}$}, is the energy
averaged over all possible values of \mbox{$X$}, given \mbox{$y$}. In this contribution, the symbol \mbox{$\mathbb{E}_{X}\{\cdot\}$} denotes expectation of the
random object within the brackets with respect to the subscript
random variable. The posterior
information (Shannon) entropy, \mbox{$H(X|Y;\beta)$}, (in nats) of the
channel can be expressed via the thermodynamic entropy conditioned
on \mbox{$Y=y$}, \mbox{$S(X|Y=y;\beta)$}, as \BEA\label{eq_S_XgivenY}
H(X|Y;\beta)&=&\mathbb{E}_{Y}\{S(X|Y=y;\beta)\}\nonumber\\
&=&-\mathbb{E}_{Y}\Bigg\{\int_{\beta}^{\infty}\gamma\frac{dU(Y;\gamma)}{d\gamma}d\gamma\Bigg\}\nonumber.
\EEA The input's entropy can also be reformulated in a similar
manner, since \mbox{$H(X)=H(X|Y;\beta=0$}). Hence, \BEA
H(X)&=&-\mathbb{E}_{Y}\Bigg\{\int_{0}^{\infty}\gamma\frac{dU(Y;\gamma)}{d\gamma}d\gamma\Bigg\}.\nonumber
\EEA Now, the input-output mutual information can be described via
thermodynamic quantities, namely the energy, \mbox{$\mathcal{E}$},
and inverse temperature, \mbox{$\beta$}, as \BEA\label{eq_I}
&&I(X;Y)=I(\beta)\triangleq H(X)-H(X|Y;\beta)\nonumber\\&&\label{eq_rhs}=-\mathbb{E}_{Y}\Bigg\{\int_{0}^{\beta}\gamma\frac{dU(Y;\gamma)}{d\gamma}d\gamma\Bigg\}\\
&&=-\big[\gamma\mathbb{E}_{Y}\{U(Y;\gamma)\}\big]_{0}^{\beta}+\mathbb{E}_{Y}\Bigg\{\int_{0}^{\beta}U(Y;\gamma)d\gamma\Bigg\}\nonumber.
\EEA For the generalized second thermodynamic law one has to insert
the effective heat contribution to the entropy and the mutual
information is given by

\BEA\label{eq_I2} &&I(X;Y)=
-\big[\gamma\mathbb{E}_{Y}\{U(Y;\gamma)\}\big]_{0}^{\beta}+\\
&&\mathbb{E}_{Y}\Bigg\{\int_{0}^{\beta}\Big(U(Y;\gamma)+
\gamma\mathbb{E}_{X|Y}\bigg\{\fd{\mathcal{E}(X|Y;\gamma)}{\gamma}\bigg\}\Big)d\gamma\Bigg\}.\nonumber
\EEA Note that this thermodynamic
expression for the mutual information holds
for any channel which can be described by a thermal system
exhibiting quasi-static heat transfer.

For the Gaussian channel with a standard Gaussian input,
\mbox{$\mathcal{N}(0,1)$}, we get
\mbox{$\log(P(X))/\beta=-x^{2}/(2\beta)$} and the
energy~(\ref{eq_energy_map}) of the Gaussian channel system becomes
an explicit function of \mbox{$\beta$}, given by
\mbox{$\mathcal{E}(X=x|Y=y;\beta)=-xy+x^2(1+\beta)/2\beta$}. The
derivative of this function with respect to \mbox{$\beta$} yields
\mbox{$d\mathcal{E}(X=x|Y=y;\beta)/d\beta=-x^{2}/(2\beta^{2})$}. The
a-posteriori probability density function is
\mbox{$p(X=x|Y=y;\beta)=\mathcal{N}(\beta
y/(1+\beta),1/(1+\beta))$.} Hence, the internal energy is \BE
U(Y=y;\beta)=\mathbb{E}_{X|Y}\{\mathcal{E}(X|Y=y;\beta)\}=-\frac{y^{2}\beta}{2(1+\beta)}+\frac{1}{2\beta}
.\nonumber\EE The derivative of the energy averaged over all
possible inputs is \BE
\mathbb{E}_{X|Y}\Bigg\{\pd{\mathcal{E}(X|Y=y;\beta)}{\beta}\Bigg\}=-\frac{1}{2\beta^{2}}\bigg(\frac{1}{1+\beta}+\frac{\beta^{2}y^{2}}{(1+\beta)^{2}}\bigg),
\nonumber\EE and the marginal probability density function of the
output is given by \mbox{$p(Y=y)=\mathcal{N}(0,(1+\beta)/\beta)$}.
Incorporating all these results into the mutual information
expression~(\ref{eq_I2}) one can easily show that \BE
I(X;Y)=\frac{1}{2}\log{(1+\beta)},\nonumber\EE and the celebrated
formula for the Shannon capacity~\cite{BibDB:Shannon} is derived
from the perspective of thermodynamics.


For the Gaussian channel with Bernoulli-\mbox{$1/2$} input, \ie~
\mbox{$P(X=1)=P(X=-1)=1/2$}, the \mbox{$X^{2}/2=1/2$} and
\mbox{$\log P(X=x)/\beta$} terms of the Gaussian channel's energy
(\ref{eq_energy_map}) are independent of \mbox{$X$} and can be
canceled out by the same terms coming from the partition function,
leaving us with the expression, independent of \mbox{$\beta$},
\mbox{$\mathcal{E}(X=x|Y=y)=-xy$}. The internal energy is
\mbox{$U(Y=y;\beta)=\mathbb{E}_{X|Y}\{\mathcal{E}(X|Y=y)\}=-y\tanh(\beta
y)$} and the marginal probability density function of the output is
then given by $p(Y=y)=(\sqrt{\beta}/(2\sqrt{2\pi}))\lbrack
exp(-\beta(y-1)^{2}/2)+exp(-\beta(y+1)^{2}/2)\rbrack$. Incorporating
these definitions into eq.~(\ref{eq_rhs}) for the case of energy
function which is independent of the temperature one can verify that
\BE
I(\beta)=\beta-\frac{1}{\sqrt{2\pi}}\int_{-\infty}^{\infty}\exp{\bigg(-\frac{y^{2
}}{2}\bigg)}\log\cosh{(\beta-\sqrt{\beta}y)}dy,\nonumber \EE which
is identical to the known Shannon-theoretic result (see,
\eg,~\cite[eq. (18)]{BibDB:GSV} and~\cite[p.
274]{BibDB:BookBlahut}).

We turn now to a discussion on the physical implication of
the generalized second thermodynamic law and the redefinition of
the temperature.

At equilibrium, a system with a $T$-dependent Hamiltonian is
occupying the macroscopic state which maximizes the entropy. We
argue that such generalized systems obey the following traditional
physical picture: (a) Two systems which  are in thermal contact and
at equilibrium have the same (redefined) temperature. (b) For two
systems which are in thermal contact and at generalized temperature
\mbox{$T_1$} and \mbox{$T_2$}, heat always flows from high
temperature to low temperature. In addition to this traditional
picture, (c) systems fulfilling \mbox{$<d\mathcal{E}/dT>~ \ge 0$}
achieve an equilibrium state, whereas systems obeying
\mbox{$<d\mathcal{E}/dT>~<0$} are unstable to thermal fluctuations,
hence cannot describe a physical system at equilibrium.

For generalized systems, the redefined inverse temperature is
\mbox{$1/T=dS/\lbrack dQ-<d\mathcal{E}/dT>dT\rbrack$}. Hence,
\mbox{$\beta$} has the same meaning as for the classical second law
but with an effective heat contributing to the entropy. When two
systems are in thermal contact the total change in the entropy is
\mbox{$dS=dS_1+dS_2$}. A necessary condition for a maximum of the
total entropy, \mbox{$dS$}, is that the slope of the entropy with an
effective heat is the same for both systems. Otherwise, heat will
flow from a system with the lower slope to the one with the higher
slope and the total entropy will increase. Hence it is clear that at
a thermal equilibrium the redefined temperatures of the two systems
have to obey \mbox{$T_1=T_2$} as for the conventional thermodynamic
picture.

We present now a self-consistent argument to prove that a
generalized system is unstable in the event of
\mbox{$<d\mathcal{E}/dT>~<0$}. Assume a generalized system is in
thermal equilibrium with a reservoir, and time-dependent
fluctuations (of heat transfers) between the measured system and the
reservoir. Fluctuations that increase (decrease) the internal energy
of the generalized system move the energy comb to `right' (`left'),
such that a higher (lower) energy is assigned to each one of the
teeth of the comb. For any macroscopic state there is a degeneracy,
\ie, many microscopic states obeying the same macroscopic
parameters. We assume that the degeneracy of the system increases in
a move to the right tooth, corresponding to a relatively higher
energy level.

For the simplicity of the discussion below, assume that the
generalized system is always occupying one tooth of the energy comb
only. In an event where the system absorbs some heat from the
reservoir by fluctuations, there are two `options' for the system
how to `invest' the absorbing heat. The first option is to move the
energy comb to the `right', and the second one is to keep the
current position of the energy comb, but to move the system to a
higher (right position) tooth. The first option does not change the
entropy, while the second one increases the entropy. Hence, the
system adopts the second option. Similarly when the system omits
heat to the reservoir the system prefers to move the energy comb to
the `left', instead of moving to a lower tooth. Such processes will
repeat again and again, until the energy comb saturates the leftmost
position, and the system moves to the highest-energy tooth,
independent of the temperature of the reservoir. When the `left'
move of the energy comb is unbounded, the process will not
terminate. Note that indeed for the Gaussian channel (see eq. 2)
\mbox{$<d\mathcal{E}/dT>=<x^2/2> ~> 0$}, and the system achieves an
equilibrium state.

Let us examine now two systems at \mbox{$T_1<T_2$},
\mbox{$<d\mathcal{E}_1/dT_1>~>0$} and
\mbox{$<d\mathcal{E}_2/dT_2>~>0$}. Assume self-consistently that the
first (second) system absorbs (emits) heat. The total change in the
entropy can be expressed as \vspace{-0.3cm}
\BEA \label{T1T2} &0&<dS=dS_1+dS_2=\frac{dQ}{T_1}\big(1-\frac{T_1}{T_2}\big)\nonumber\\
&-&<\frac{d\mathcal{E}_1}{dT_1}>\frac{dT_1}{T_1}\Big\lbrack
1-\big(\frac{
<\frac{d\mathcal{E}_2}{dT_2}>}{<\frac{d\mathcal{E}_1}{dT_1}>}\big)\big(\frac{T_1}{T_2}\big)\big(\frac{dT_2}{dT_1}\big)\Big\rbrack,
\EEA where it is clear that \mbox{$dT_1>0$} and \mbox{$dT_2<0$}. (If
\mbox{$dT_1<0$} then the first system absorbs more heat than
\mbox{$dQ$}, \mbox{\ie,} \mbox{$dQ-<d\mathcal{E}_1/dT_1>dT_1$}.) Now
it can be easily verified that the last term (second line) of the
r.h.s of ~(\ref{T1T2}) is negative. Hence, a necessary condition for
the total entropy, \mbox{$dS$}, to be positive is that the first
term of the r.h.s. of ~(\ref{T1T2}) is positive, \mbox{$T_2>T_1$},
implying that heat flows from high to low temperatures.

$T$-dependent (independent) Hamiltonians arise also for discrete
(continuous) input symbols. An example for such \mbox{$T$}-dependent
Hamiltonians is the Gaussian channels with non-equiprobable binary
inputs, \mbox{$P(\pm1)\ne1/2$}, whereas an example for a
\mbox{$T$}-independent Hamiltonian is the Gaussian channel with a
uniform input distribution. It is clear that the characterization of
communication channels with \mbox{$T$}-dependent/independent
Hamiltonians does not follow the type of the input/output symbols.
What is the origin, from the point of  view of physics, for the
\mbox{$T$}-dependent Hamiltonians? In order to answer this question
one has to examine a possible conflict between the nature of the
heat and the constraint imposed by the communication channel.

For a communication channel governed by a \mbox{$T$}-dependent
Hamiltonian, \eg~eq.~(2),  the distribution of a microscopic state
\mbox{$exp(-\beta H) \propto P(X)$}  which is independent of
\mbox{$\beta$}. This is a nonphysical postulate assigned  to the
hidden values of the input \mbox{$X$}, a degree of freedom in the
physical system. The distribution of a degree of freedom (\eg, spin)
governed by a Hamiltonian system with/without an inversion symmetry
can be nonuniform, but it  always depends on the temperature. There
is a natural distribution inspired by the heat and a distribution
imposed by the communication channel. In case of a conflict between
these two distributions the constraint imposed by the channel wins,
and the system has to devote some heat to implement it. A heat
`piston' is required to bend the natural heat-distribution to that
of the channel, and this is the origin for the {\it working channel}
by the communication-heat-engine (Figure 1).

The non-prefect communication-heat-engine has also a meaningful
fingerprint in the additional (third term) of the generalized mutual
information, eq. (6), which is negative  since
\mbox{$<d\mathcal{E}/d\beta>~<0$}, (\mbox{$<d\mathcal{E}/dT>~>0$}). It indicates
that the waste of heat playing against the laws of nature decreases
the mutual information in comparison to the same model
(communication channel) with the lack of the \mbox{$T$}-dependent term in
the Hamiltonian.

\comment{


}

\end{document}